\documentclass{emulateapj}

\newcommand{\kms}{km~s$^{-1}$}
\newcommand{\msun}{M$_{\odot}$}
\newcommand{\myr}{M$_{\odot}$~yr$^{-1}$}
\newcommand{\ngal}{1315}

\shorttitle{Mid-Infrared Butcher-Oemler Effect}
\shortauthors{Saintonge, Tran \& Holden}

\begin{document}

\title{{\it Spitzer}/MIPS 24 $\mu$\MakeLowercase{m} Observations of Galaxy Clusters: \\
An Increasing Fraction of Obscured Star-forming Members from \MakeLowercase{z}=0.02 to \MakeLowercase{z}=0.83 }

\author{Am\'{e}lie Saintonge\altaffilmark{1}, Kim-Vy H. Tran\altaffilmark{1} and Bradford P. Holden\altaffilmark{2}}
\altaffiltext{1}{Institute for Theoretical Physics, University of Zurich, CH-8057 Zurich, Switzerland}
\email{amelie@physik.uzh.ch}
\altaffiltext{2}{UCO/Lick Observatories, University of California, Santa Cruz, CA 95064}

\begin{abstract}

We study the mid-infrared properties of \ngal~ spectroscopically
confirmed members in eight massive ($M_{vir}\gtrsim5\times10^{14}$\msun) galaxy clusters covering the redshift range from 0.02 to 0.83.
The selected clusters all have deep {\it Spitzer} MIPS $24\mu$m
observations, {\it Hubble} and ground-based photometry, and extensive
redshift catalogs.  We observe for the first time an increase in the fraction of cluster
galaxies with mid-infrared star formation rates higher than 5 \myr~
from 3\% at $z=0.02$ to 13\% at $z=0.83$ ($R_P\leqslant 1$Mpc).  This increase is reproduced even 
when considering only the most massive members ($M_{\ast} \geqslant 4 \times 10^{10}$\msun).
The $24\mu$m observations reveal stronger evolution in the fraction of blue/star-forming
cluster galaxies than color-selected samples: 
the number of dusty, strongly star-forming cluster galaxies increases
with redshift, and combining these with the optically-defined
Butcher-Oemler members [$\Delta(B-V)<-0.2$] doubles the total
fraction of blue/star-forming galaxies in the inner Mpc of the clusters to $\sim23$\% at
$z=0.83$.  These results, the first of our {\it Spitzer}/MIPS Infra-Red Cluster Survey (SMIRCS), support earlier studies indicating the
increase in star-forming members is driven by cluster assembly and galaxy infall,
as is expected in the framework of hierarchical formation.

\end{abstract}


\section{Introduction}

\citet{bo78,bo84} observed that galaxy clusters at intermediate
redshift have a higher fraction of members with blue optical colors
than clusters in the local universe, thus providing a key piece of
evidence supporting galaxy evolution.  This increase in blue members
with redshift, named the Butcher-Oemler (BO) effect, was intensely
debated for two decades \citep[e.g.][]{mathieu81,dressler82}. However, multiple optical
studies based on spectroscopic observations have since confirmed the
increase in blue, star-forming galaxies in higher redshift clusters
\citep[e.g.][]{couch87,caldwell97,fisher98,ellingson01}, and
found that BO galaxies reveal signs of recent and
ongoing star formation.  The paramount question now is have we seen
only the tip of the iceberg?

Most studies of star-forming galaxies in clusters rely on rest-frame
ultraviolet or optical tracers \citep[e.g.][]{balogh98,poggianti06},
but UV/optical tracers can suffer severely from dust obscuration,
especially when star formation is concentrated in the nuclear regions
\citep{kennicutt98}.  For example, ultraluminous infrared galaxies
have SF rates of $\gtrsim1000$\msun, yet many ULIRGs fail to even be
detected at UV and optical wavelengths \citep[e.g.][]{houck05}. Although
corrections for dust attenuation are possible, reliable estimates of
SF rates cannot be achieved solely using rest-frame UV/optical
observations \citep{bell02,cardiel03}.

A substantially more robust method of determining total SF rates is
with mid-infrared (MIR) imaging.  The first MIR imaging of galaxy
clusters at intermediate redshifts was taken with ISO's ISOCAM camera, 
and \citet{duc02} found that at least 90\% of the
star formation was hidden at optical wavelengths.  The first handful
of galaxy clusters observed with the MIPS camera on the {\it Spitzer
Space Telescope} (SST) have also revealed strong dust-obscured star
formation \citep{geach06,marcillac07,bai07}.

It remains unclear as to what causes the increase in star-forming 
galaxy cluster members.  Detailed
morphological studies of blue galaxies [defined as having
$\Delta(B-V)<-0.2$]\footnote{$\Delta(B-V)$ is the color offset from the red 
sequence fit to the cluster ellipticals.} 
with the {\it Hubble Space Telescope} (HST) find
that most are disk systems similar to those in local clusters
\citep[e.g.][]{dressler94,couch94}; past studies also find that many
show signs of interactions or mergers
\citep{lavery88,lavery92,couch94,oemler97}.  More recently, studies
indicate that galaxy infall is a viable explanation for the
significant numbers of blue galaxies and their disturbed morphologies
in intermediate redshift clusters
\citep[e.g.][]{vandokkum98,ellingson01,tran05}, a scenario supported
by hierarchical clustering models \citep{kauffmann95}.  In this case,
galaxy clusters that are accreting a significant number of new members
should have a higher fraction of star-forming galaxies, especially at 
higher redshifts when the amount of activity was enhanced  
also in the field. 

Here we present the first comprehensive study of SST/MIPS $24\mu$m
imaging of galaxies that are spectroscopically confirmed members of
eight massive ($M_{vir}\gtrsim5\times10^{14}$\msun) X-ray luminous
clusters spanning a wide redshift range ($0.02<z<0.83$). After
presenting the data in \S2, we focus
our analysis in \S3 and \S4 on the evolution of star-forming members
with redshift.  A
cosmology with $(H_0,\Omega_{M},\Omega_{\Lambda})=(70$ \kms$,
0.3,0.7)$ is assumed throughout the paper; at $z=0.83$, the look-back
time is $\sim7$ Gyr.\\

\section{Data}

\begin{deluxetable*}{lccccccccc}
\tabletypesize{\scriptsize}
\tablecaption{Properties of selected clusters \label{clusters}}
\tablehead{
\colhead{Name} & \colhead{Coords [J2000]}  & \colhead{$z$ range\tablenotemark{a}} & \colhead{$L_X$\tablenotemark{b}} &
\colhead{$N_{z}$\tablenotemark{c}} & \colhead{$N_s$\tablenotemark{d}} & 
\colhead{$N_{24}$\tablenotemark{e}} & \colhead{$t_{int}$}  &
\colhead{$F_{bg}$}  & \colhead{$F_{50\%}$}\\
&  &  & ($10^{44}$ erg s$^{-1}$) &  &  &  &   \colhead{(s pix$^{-1}$)}  & \colhead{(MJy sr$^{-1}$)} & \colhead{(\myr)}
}
\startdata
Coma & 125935.7+275734       &  0.013-0.033  & $9.0\pm0.2$ &244 &63  &134 (2) &73 &32.8 &0.02 \\
Abell 1689 & 131129.5-012017 & 0.17-0.22 & $21.4\pm1.0$ & 81  &52   &12 (2)   &\nodata\tablenotemark{f}  &\nodata &1.4\tablenotemark{f} \\
MS 1358+62 & 135950.4+623103   &  0.315-0.342  & $10.2\pm0.7$  &171 &73   &21 (3)   &2700\tablenotemark{g} &20.6 &0.75\\
CL 0024+17 & 002635.7+170943    &  0.373-0.402  & $2.9\pm0.1$  &205 &51  &11 (6)    &2700\tablenotemark{g} &48.5 &1.48\\
MS 0451--03 & 045410.9-030107    & 0.52-0.56  & $21.0\pm0.4$  &242  &38   &8 (5)      &2700\tablenotemark{g} &35.0 &3.34\\
MS 2053--04 & 205621.3-043751   & 0.57-0.60  & $6.5\pm0.4$  &85    &43   &15 (8)    &1950 &35.2  &5.04\\
MS 1054--03 & 105700.0-033736    & 0.80-0.86  & $16.4\pm0.8$  &142  &75   &13 (8)    &3600\tablenotemark{g} &47.4 &4.54\\
RX J0152--13 & 015243.9-135719  & 0.81-0.87 & $18.6\pm1.9$   &147  &61   &19 (8)    & 3600\tablenotemark{g} &31.9  &3.05\\
\enddata
\tablenotetext{a}{Cluster members selected within this redshift range, as in H07 (Coma, MS1358, MS2053, MS1054, RXJ0152), \citet{duc02} (A1689) and \citet{moran07} (CL0024 and MS0451).}
\tablenotetext{b}{Bolometric X-ray luminosities from H07 (Coma, MS1358, MS2053, MS1054, RXJ0152), \citet{bardeau07} (Abell1689), \citet{donahue99} (MS0451),  \citet{zhang05} (CL0024).}
\tablenotetext{c}{Total number of spectroscopically confirmed members
(magnitude-limited selection). Redshifts are taken from \citet{beijersbergen03},\citet{duc02},\citet{fisher98},\citet{moran05},
\citet{moran07},\citet{tran05},\citet{tran07},\citet{demarco05}, respectively.} 
\tablenotetext{d}{Number of confirmed members within 1 Mpc of the
cluster center and brighter than $M_B=-19.5+5\log h$.}
\tablenotetext{e}{Number of MIPS detections in the cluster; () is the
number of galaxies within $N_s$ with SF rates $\geqslant5$ \myr.}
\tablenotetext{f}{We are using ISOCAM mid-IR data from \citet{duc02} for A1689.}
\tablenotetext{g}{Over the central 5\arcmin$\times$5\arcmin of the MIPS image.}
\end{deluxetable*}

We have assembled a data set of eight galaxy clusters at $0.02\leq
z\leq0.83$ that have a total of \ngal~ spectroscopically confirmed
members.  The core of our sample is composed of five
clusters spanning the entire redshift range with large spectroscopic
membership, uniform multi-filter optical photometry and deep 
SST/MIPS imaging\footnote{This work is based on observations made with the
Spitzer Space Telescope, which is operated by the Jet Propulsion
Laboratory, California Institute of Technology under a contract with
NASA.}.  For the part
of the analysis that does not depend on rest-frame $(B-V)$ color, we
fold into the sample three additional clusters: Abell 1689 for which
MIR data from ISOCAM is available \citep{duc02}, and CL0024 and
MS0451, both of which have extensive redshift catalogs \citep{moran05}
and MIPS observations. Observational details for all
clusters are in Table~\ref{clusters}.

\subsection{Optical Photometry and Spectroscopy \label{optical}}

The optical photometry for the five main clusters is from
\citet[][hereafter H07]{holden07} where magnitudes and
colors were derived from Sersic models fitted to HST/WFPC2 images
(MS1358, MS2053, and RXJ0152), HST/ACS images (MS1054), and SDSS
mosaics for Coma.  The conversion to rest-frame values is done by 
interpolating between the passbands \citep{blakeslee06} and has errors of $\sim0.02$ mag. 
The mass-to-light ratios ($M/L_{B}$) and
stellar masses were calculated using the relation between rest-frame
$(B-V)$ color and $M/L_{B}$; see H07 for details and a discussion on
the associated errors.

\subsection{MIPS $24\mu$m Imaging \label{mips}}

All MIPS data sets were retrieved from the $Spitzer$ public archive. 
Individual frames were corrected with scan mirror
position-dependent flats and then mosaiced with the MOPEX software \citep{mopex05} 
to a pixel size of 1.2\arcsec \footnote{The instrumental pixels are 2.55\arcsec in size, but the 
finer sampling helps in 
improving the characterization of the PSF.}. Integration times ($t_{int}$) and background levels ($F_{bg}$) in
these mosaics are given in Table \ref{clusters}.  Photometry was
performed with APEX \citep{apex05} using a 3\arcsec -diameter aperture, and
an aperture correction of 9.49 as given in the MIPS data handbook. 
A small aperture is necessary to avoid contamination in the deep and crowded 
cluster fields. The fluxes are consistent with results from PSF-fitting photometry 
with scatter from a 1:1 relation in the range of 15-25 $\mu$Jy.

To estimate the completeness of each MIPS catalog, we 
added to the mosaics artificial sources modeled on the PSF. To
avoid overcrowding, we simulated 30 signals at once, and repeated the
process 30 times for each cluster (the $50\%$ completeness limits, $F_{50\%}$, are 
presented in Table \ref{clusters}). Finally, the MIPS sources were matched 
with the optical catalogs using a 2\arcsec~search radius \citep{bai07}. From randomization of 
the MIPS coordinates, we estimate the rate of false identification to be $7\pm4\%$, 
and little dependency of this error rate on redshift or color is observed.

\subsubsection{Star formation rates}

Star formation rates are based on the $24\mu$m fluxes. First, the
total infrared luminosity ($F_{8-1000\mu {\rm m}}$) of each galaxy was
determined using a family of infrared spectral energy distributions
(SEDs) from \citet{dale02}. We choose the range of SEDs that are
representative of the galaxies in the Spitzer Infrared Nearby Galaxies
Survey \citep{dale07}, and at each redshift adopt the median
conversion factor from $F_{24\mu{\rm m}}$ to $F_{8-1000\mu{\rm m}}$
given by these models. At $0.4\lesssim z\lesssim 0.6$, the error due
to the adopted conversion factor is $\sim20$\%, but the error
increases to a factor of 1.5-2.0 at lower and higher redshifts.  For
the parts of our analysis that are sensitive to the SF rates, we take
these errors into account.  As a check, we note that our total
infrared luminosities in MS1054 agree well with the values in
\citet{bai07}. The conversion from total infrared luminosities to 
star formation rates is done following \citet{kennicutt98}.

We assume that the emission at $24\mu{\rm m}$ is due to star formation
but it could also be due to dust-enshrouded active galactic nuclei
(AGN).  However, in comparing the X-ray and $24\mu{\rm m}$ detections,
only one cluster galaxy (in RXJ0152) is detected in both and rejected.  While the
AGN fraction in clusters seems to increase with redshift
\citep{eastman07}, the estimated AGN fraction is only 2\% at
$z\sim0.6$.  \citet{johnson03} also find evidence that at $z\sim0.8$,
any excess X-ray AGN are located at $R>1$~Mpc whereas we focus on the
central Mpc of each cluster.  Although we cannot completely rule out
possible contamination by weak obscured AGN, we have excluded X-ray
AGN and thus assume that the galaxies detected by MIPS are powered by
dusty star formation; see \citet{marcillac07} for a more detailed
argument on why this is a reasonable assumption.

\section{Results}

\subsection{Color-Magnitude diagrams}

\begin{figure}
\epsscale{1.0}
\plotone{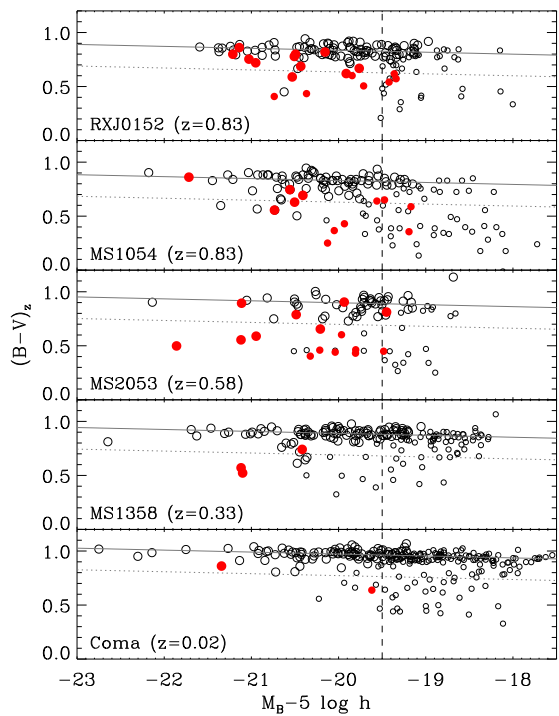}
\caption{Rest-frame color-magnitude diagrams for
spectroscopically confirmed members in the main cluster sample. Filled
red circles are MIPS detections with SF rates $\geqslant5$ \myr (where all clusters 
are better than 50\% complete).  The larger symbols represent galaxies with 
$\log_{10}({\rm M}_{\ast}) \geqslant 10.6$.
The rest-frame B-band magnitude has been corrected for passive luminosity evolution, as
determined from the fundamental plane \citep{vandokkum98L}.  The vertical
dashed line is the rest-frame B-band magnitude selection limit of
-19.5. The solid diagonal line is the best fit to the red sequence
galaxies, adopting the slope of \citet{vandokkum98}, and the dotted
line denotes $\Delta(B-V)=-0.2$ mag; only galaxies below the dotted
line would be part of standard BO sample. \label{fig1}}
\end{figure}

Figure~\ref{fig1} presents the color-magnitude diagrams of the
five main clusters with photometry from H07.  Because the MIPS 
sensitivity varies from cluster to cluster, we apply a SF
rate limit of 5 \myr.  The first immediate observation is that
the number of strongly star-forming galaxies increases significantly
with redshift.  Using a field galaxy sample drawn from the same 
photometric and spectroscopic catalogs, we estimate a possible field 
contamination at $z=0.83$ to be $\sim8$\%  (i.e. no more than one galaxy per cluster).
In Figure~\ref{fig1}, the dotted
lines represent the original color criterion for BO galaxies. The
ratio of the number of cluster galaxies with MIR SF rate $\geqslant5$\myr~ above
this color cut to the number of blue galaxies ($\Delta(B-V)<-0.2$) 
increases with redshift. 

\subsection{The Mid-Infrared Butcher-Oemler effect}

\begin{figure}
\epsscale{1.0}
\plotone{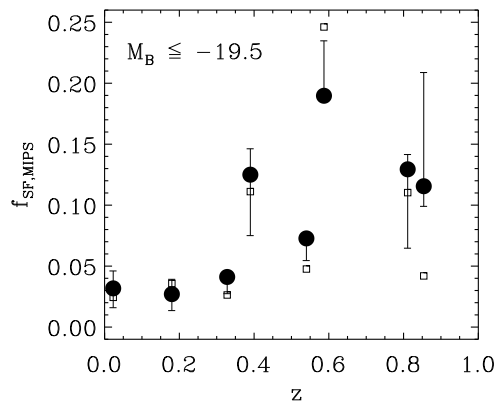}
\caption{Fraction of confirmed cluster galaxies that are star-forming
as revealed by the MIPS $24\mu$m observations. Are considered 
only members with MIR SF rates
$\geqslant5$ \myr~ that are brighter than $M_B=-19.5$ and located
within 1 Mpc of the cluster centers (filled circles) and 500 kpc (open squares).  
The points for the two $z\sim0.83$ clusters
are offset slightly in $z$ for clarity.\label{fig2}}
\end{figure}

\begin{figure}
\epsscale{1.0}
\plotone{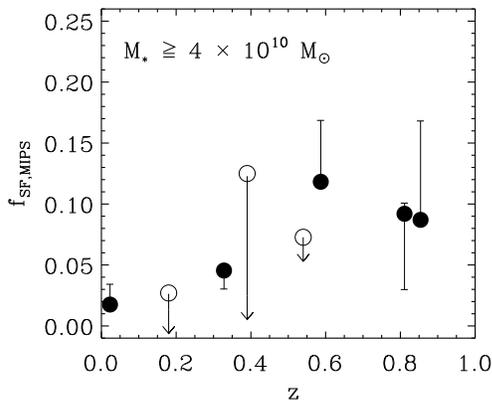}
\caption{As in Fig.~\ref{fig2}, the fraction of confirmed star-forming
cluster galaxies (MIR SF rate$\geqslant5$\myr, $M_B\leq-19.5$, $R<1$
Mpc), but now with the additional stellar mass cut-off of $\log_{10}({\rm
M}_{\ast}) \geqslant 10.6$ for the five main clusters (filled
circles).  Stellar masses are not available for the remaining three
clusters; they are  shown as upper limits (open circles).\label{fig3}} 
\end{figure}

For each cluster, we compute and plot in Figure \ref{fig2} the
fraction of confirmed star-forming cluster members after selecting by 
rest-frame B-band magnitude ($M_B\leq-19.5$), cluster-centric
distance\footnote{While the optical 
observations generally extend to $R>1.5$ Mpc, the MIPS imaging for MS1358 only extends to $\sim1$
Mpc ($\sim50-60$\% of $r200$ for these clusters).}, and MIR star formation rate ($\geqslant5$\myr).  
The errors on $f_{SF,MIPS}$ represent the range that can 
be produced by taking the minimum and maximum conversion 
factors from $F_{24\mu{\rm m}}$ to $F_{8-1000\mu{\rm
m}}$ instead of a single average value for each cluster, and by 
varying the different selection thresholds by amounts comparable 
to the errors on each of these parameters.

Figure \ref{fig2} shows that the fraction of galaxies in clusters with
MIR SF rates $\geqslant5$\myr~ steadily climbs from $\sim3$\% locally
to $\sim13$\% at $z=0.83$.  Because H07 showed that a cluster's
morphological composition can vary depending on whether members are
selected by mass or by luminosity, we apply an additional stellar mass
cut of $\log_{10}({\rm M}_{\ast}) \geqslant 10.6$ (Fig.~\ref{fig3}).
The mass cut is only applied to the five main clusters for which
uniform photometry and thus stellar masses are available; 
the remaining three clusters are shown only as upper limits.  While the mass cut
attenuates the increase in fraction of star-forming members, it does
not completely suppress the trend. Thus the MIR BO effect 
is not due to an increase in the fraction of
faint, low-mass members temporarily brightened by strong star
formation.

\section{Discussion}

Having established an increase in the fraction of MIR-detected galaxies from $z\sim0$ to
$z\sim0.8$, we stress that optical studies are likely underestimating
the increase in star-forming cluster galaxies with redshift.  As seen
in Fig.~\ref{fig1}, an increasing number of strong dust-obscured
star-forming members appear on or near the red sequence at higher redshifts;
these are not included in traditional color-selected BO studies.  The late-type
morphologies of these members supports our intepretation of dusty star
formation and red colors due to extinction \citep[see also
A901/902;][]{wolf05}.  

Using the standard BO definition of
$\Delta(B-V)<-0.2$, the fraction of blue galaxies with $M_B\leq-19.5$ 
and $R_P<1$~Mpc at $z\sim0.8$ is
$\sim11$\%; however, including the red, massive, star-forming members raises
the total fraction of blue/star-forming members to $\sim23$\%.  We
note that for the five main clusters, the increase in the
blue/star-forming fraction due to these red, star-forming members is
\{1.1, 1.2, 1.3, 1.7\} at $z=$\{0.02, 0.33, 0.59, 0.83\}, $i.e.$ the
relative importance of including red, dusty star-forming members
increases with redshift.

Is this increase linked to galaxy infall?  In Figure~\ref{fig2}, both
CL0024 ($z\sim0.4$) and MS2053 ($z\sim0.6$) are above the general
trend established by the other six clusters.  Both CL0024 and MS2053
have enhanced star formation compared to other clusters at similar
redshift, and both have bimodal redshift distributions.  CL0024 is made of two colliding
subclusters \citep{czoske02}, and has an unusually large number of
luminous infrared galaxies \citep{coia05}.  Similarly, \citet{tran05}
conclude from that MS2053 has a
significant number ($>25$\%) of infalling galaxies; these members tend
to be blue and star-forming. Both CL0024 and MS2053 are accreting a
large number of new members and have high fractions of dusty
star-forming galaxies.  We speculate that the increase in star-forming
members reflects the recent accretion of new members, $i.e.$ galaxy
infall, and that such events are more frequent at higher redshift due
to the process of cluster assembly \citep{ellingson01,tran05,loh08}. 
As further evidence of this, 80\% of the MIPS-detected galaxies in the 
$z\sim0.8$ clusters are more than 700 kpc from the cluster cores 
in projected distance and thus the MIR Butcher-Oemler effect  
is significantly altered by only considering the inner 500 kpc of the clusters
(open symbols in Fig.\ref{fig2}).

\section{Summary}

We present the first comprehensive study of SST/MIPS 24$\mu$m
observations for seven massive, X-ray luminous galaxy clusters
spanning a wide redshift range ($0.02<z<0.83$).  Uniform photometry,
high resolution HST imaging, and extensive redshift catalogs enable us
to measure the fraction of members with strong, dust-obscured star
formation.  The fraction of cluster galaxies with MIR
star formation rates $\geqslant5$\myr~ increases from 3\% in Coma to
$\sim13$\% in clusters at $z=0.83$, and this trend is evident in
both luminosity ($M_B\leq-19.5$) and mass-selected samples ($M_{\ast} \geqslant 4 \times 10^{10}$\msun).

Optically-based studies increasingly underestimate the total amount of
star formation in cluster galaxies with redshift because many of these
dusty red star-forming members are missed in color-selected samples.  
These tend to be late-type galaxies that are red because of dust extinction which
disguises their high levels of obscured star formation ($>5$\myr).
Defining the SF fraction to include both optically blue and red, 
but MIPS-detected members doubles the fraction at $z=0.83$ from $\sim11$\% to
$\sim23$\% ($R_P<1$Mpc).

Lastly, our study indicates that the BO effect and the increase in 
obscured star-forming members are linked to galaxy 
infall: 80\% of the MIR-detected members at $z\sim0.8$ are outside  
the cluster cores ($R_P>0.7$Mpc), and the two clusters at $z<0.8$ that are accreting a substantial number of
new members also have an enhanced fraction of galaxies with MIR SF
rates $\geqslant5$\myr.

\acknowledgments

We are grateful to C. Papovich and L. Bai for advice on MIPS data
reduction and to G. Rudnick for useful discussions.  
AS and KT acknowledge support from 
the Swiss National Science Foundation (grant PP002-110576).

{\it Facilities:} \facility{HST (WFPC2, ACS)}, \facility{Spitzer (MIPS)}.


\begin{thebibliography}{48}
\expandafter\ifx\csname natexlab\endcsname\relax\def\natexlab#1{#1}\fi

\bibitem[{Bai} {et~al.}(2007)]{bai07}
{Bai}, L., et al. 2007, \apj,  664, 181

\bibitem[{{Balogh} {et~al.}(1998){Balogh}, {Schade}, {Morris}, {Yee},
  {Carlberg}, \& {Ellingson}}]{balogh98}
{Balogh}, M.~L., {Schade}, D., {Morris}, S.~L., {Yee}, H.~K.~C., {Carlberg},
  R.~G., \& {Ellingson}, E. 1998, \apjl, 504, L75+

\bibitem[{Bardeau} {et~al.}(2007)]{bardeau07}
{Bardeau}, S., et al. 2007, \aap, 470, 449

\bibitem[{{Beijersbergen}(2003)}]{beijersbergen03}
{Beijersbergen}, M. 2003,  Ph.D. Thesis, Rijksuniversiteit Groningen

\bibitem[{{Bell}(2002)}]{bell02}
{Bell}, E.~F. 2002, \apj, 577, 150

\bibitem[{Blakeslee} {et~al.}(2006)]{blakeslee06}
{Blakeslee}, J.~P., et al. 2006, \apj, 644, 30

\bibitem[{{Butcher} \& {Oemler}(1978)}]{bo78}
{Butcher}, H., \& {Oemler}, Jr., A. 1978, \apj, 219, 18

\bibitem[{{Butcher} \& {Oemler}(1984)}]{bo84}
---. 1984, \apj, 285, 426

\bibitem[{{Caldwell} \& {Rose}(1997)}]{caldwell97}
{Caldwell}, N., \& {Rose}, J.~A. 1997, \aj, 113, 492

\bibitem[{{Cardiel} {et~al.}(2003){Cardiel}, {Elbaz}, {Schiavon}, {Willmer},
  {Koo}, {Phillips}, \& {Gallego}}]{cardiel03}
{Cardiel}, N., {Elbaz}, D., {Schiavon}, R.~P., {Willmer}, C.~N.~A., {Koo},
  D.~C., {Phillips}, A.~C., \& {Gallego}, J. 2003, \apj, 584, 76

\bibitem[{{Coia} {et~al.}(2005){Coia}, {McBreen}, {Metcalfe}, {Biviano}, \&
  et~al.}]{coia05}
{Coia}, D., et~al. 2005, \aap,  431, 433

\bibitem[{{Couch} {et~al.}(1994){Couch}, {Ellis}, {Sharples}, \&
  {Smail}}]{couch94}
{Couch}, W.~J., {Ellis}, R.~S., {Sharples}, R.~M., \& {Smail}, I. 1994, \apj,
  430, 121

\bibitem[{{Couch} \& {Sharples}(1987)}]{couch87}
{Couch}, W.~J., \& {Sharples}, R.~M. 1987, \mnras, 229, 423

\bibitem[{{Czoske} {et~al.}(2002){Czoske}, {Moore}, {Kneib}, \&
  {Soucail}}]{czoske02}
{Czoske}, O., {Moore}, B., {Kneib}, J.-P., \& {Soucail}, G. 2002, \aap, 386, 31

\bibitem[{Dale} {et~al.}(2007)]{dale07}
{Dale}, D.~A., et~al.  2007, \apj, 655, 863

\bibitem[{{Dale} \& {Helou}(2002)}]{dale02}
{Dale}, D.~A., \& {Helou}, G. 2002, \apj, 576, 159

\bibitem[{Demarco} {et~al.}(2005)]{demarco05}
{Demarco}, R., et~al. 2005,  \aap, 432, 381

\bibitem[{{Donahue} {et~al.}(1999){Donahue}, {Voit}, {Scharf}, {Gioia},
  {Mullis}, {Hughes}, \& {Stocke}}]{donahue99}
{Donahue}, M., {Voit}, G.~M., {Scharf}, C.~A., {Gioia}, I.~M., {Mullis}, C.~R.,
  {Hughes}, J.~P., \& {Stocke}, J.~T. 1999, \apj, 527, 525

\bibitem[{{Dressler} \& {Gunn}(1982)}]{dressler82}
{Dressler}, A., \& {Gunn}, J.~E. 1982, \apj, 263, 533

\bibitem[{{Dressler} {et~al.}(1994){Dressler}, {Oemler}, {Butcher}, \&
  {Gunn}}]{dressler94}
{Dressler}, A., {Oemler}, A.~J., {Butcher}, H.~R., \& {Gunn}, J.~E. 1994, \apj,
  430, 107

\bibitem[{Duc} {et~al.}(2002)]{duc02}
{Duc}, P.-A., et al. 2002, \aap, 382, 60

\bibitem[{{Eastman} {et~al.}(2007){Eastman}, {Martini}, {Sivakoff}, {Kelson},
  {Mulchaey}, \& {Tran}}]{eastman07}
{Eastman}, J., {Martini}, P., {Sivakoff}, G., {Kelson}, D.~D., {Mulchaey},
  J.~S., \& {Tran}, K.-V. 2007, \apjl, 664, L9

\bibitem[{{Ellingson} {et~al.}(2001){Ellingson}, {Lin}, {Yee}, \&
  {Carlberg}}]{ellingson01}
{Ellingson}, E., {Lin}, H., {Yee}, H.~K.~C., \& {Carlberg}, R.~G. 2001, \apj,
  547, 609

\bibitem[{{Fisher} {et~al.}(1998){Fisher}, {Fabricant}, {Franx}, \& {van
  Dokkum}}]{fisher98}
{Fisher}, D., {Fabricant}, D., {Franx}, M., \& {van Dokkum}, P. 1998, \apj,
  498, 195

\bibitem[{Geach} {et~al.}(2006)]{geach06}
{Geach}, J.~E., et al. 2006, \apj, 649, 661

\bibitem[{Holden} {et~al.}(2007)]{holden07}{Holden}, B.~P., et~al. 2007, \apj, 670, 190

\bibitem[{{Houck} {et~al.}(2005){Houck}, {Soifer}, {Weedman}, {Higdon},
  {Higdon}, \& et~al.}]{houck05}
{Houck}, J.~R., et~al. 2005, \apjl, 622, L105

\bibitem[{{Johnson} {et~al.}(2003){Johnson}, {Best}, \& {Almaini}}]{johnson03}
{Johnson}, O., {Best}, P.~N., \& {Almaini}, O. 2003, \mnras, 343, 924

\bibitem[{{Kauffmann}(1995)}]{kauffmann95}
{Kauffmann}, G. 1995, \mnras, 274, 153

\bibitem[{{Kennicutt}(1998)}]{kennicutt98}
{Kennicutt}, Jr., R.~C. 1998, \araa, 36, 189

\bibitem[{{Lavery} \& {Henry}(1988)}]{lavery88}
---. 1988, \apj, 330, 596

\bibitem[{{Lavery} {et~al.}(1992){Lavery}, {Pierce}, \& {McClure}}]{lavery92}
{Lavery}, R.~J., {Pierce}, M.~J., \& {McClure}, R.~D. 1992, \aj, 104, 2067

\bibitem[{{Loh} {et~al.}(2008){Loh}, {Ellingson}, {Yee}, {Gilbank}, {Gladders},
  \& {Barrientos}}]{loh08}
{Loh}, Y., {Ellingson}, E., {Yee}, H.~K.~C., {Gilbank}, D.~G., {Gladders},
  M.~D., \& {Barrientos}, L.~F. 2008, \apj, 680, 214 

\bibitem[{{Makovoz} \& {Khan}(2005)}]{mopex05}
{Makovoz}, D., \& {Khan}, I. 2005, in ASP Conference Series, Vol. 347, ADASS  XIV, 
ed. P.~{Shopbell}, M.~{Britton}, \& R.~{Ebert}, 81

\bibitem[{{Makovoz} \& {Marleau}(2005)}]{apex05}
{Makovoz}, D., \& {Marleau}, F.~R. 2005, \pasp, 117, 1113

\bibitem[{{Marcillac} {et~al.}(2007){Marcillac}, {Rigby}, {Rieke}, \&
  {Kelly}}]{marcillac07}
{Marcillac}, D., {Rigby}, J.~R., {Rieke}, G.~H., \& {Kelly}, D.~M. 2007, \apj,
  654, 825

\bibitem[{{Mathieu} \& {Spinrad}(1981)}]{mathieu81}
{Mathieu}, R.~D., \& {Spinrad}, H. 1981, \apj, 251, 485

\bibitem[{{Moran} {et~al.}(2005){Moran}, {Ellis}, {Treu}, {Smail}, {Dressler},
  {Coil}, \& {Smith}}]{moran05}
{Moran}, S.~M., {Ellis}, R.~S., {Treu}, T., {Smail}, I., {Dressler}, A.,
  {Coil}, A.~L., \& {Smith}, G.~P. 2005, \apj, 634, 977

\bibitem[{{Moran} {et~al.}(2007){Moran}, {Ellis}, {Treu}, {Smith}, {Rich}, \&
  {Smail}}]{moran07}
{Moran}, S.~M., {Ellis}, R.~S., {Treu}, T., {Smith}, G.~P., {Rich}, R.~M., \&
  {Smail}, I. 2007, \apj, 671, 1503

\bibitem[{{Oemler} {et~al.}(1997){Oemler}, {Dressler}, \& {Butcher}}]{oemler97}
{Oemler}, A.~J., {Dressler}, A., \& {Butcher}, H.~R. 1997, \apj, 474, 561

\bibitem[{Poggianti} {et~al.}(2006)]{poggianti06}
{Poggianti}, B.~M., et~al. 2006,  \apj, 642, 188

\bibitem[{Tran} {et~al.}(2007)]{tran07}
{Tran}, K.-V.~H., et al. 2007, \apj, 661, 750

\bibitem[{{Tran} {et~al.}(2005){Tran}, {van Dokkum}, {Illingworth}, {Kelson},
  {Gonzalez}, \& {Franx}}]{tran05}
{Tran}, K.-V.~H., {van Dokkum}, P., {Illingworth}, G.~D., {Kelson}, D.,
  {Gonzalez}, A., \& {Franx}, M. 2005, \apj, 619, 134

\bibitem[{{van Dokkum} {et~al.}(1998{\natexlab{a}}){van Dokkum}, {Franx},
  {Kelson}, \& {Illingworth}}]{vandokkum98L}
{van Dokkum}, P.~G., {Franx}, M., {Kelson}, D.~D., \& {Illingworth}, G.~D.
  1998{\natexlab{a}}, \apjl, 504, L17+

\bibitem[{{van Dokkum} {et~al.}(1998{\natexlab{b}}){van Dokkum}, {Franx},
  {Kelson}, {Illingworth}, {Fisher}, \& {Fabricant}}]{vandokkum98}
{van Dokkum}, P.~G., {Franx}, M., {Kelson}, D.~D., {Illingworth}, G.~D.,
  {Fisher}, D., \& {Fabricant}, D. 1998{\natexlab{b}}, \apj, 500, 714

\bibitem[{{Wolf} {et~al.}(2005){Wolf}, {Gray}, \& {Meisenheimer}}]{wolf05}
{Wolf}, C., {Gray}, M.~E., \& {Meisenheimer}, K. 2005, \aap, 443, 435

\bibitem[{{Zhang} {et~al.}(2005){Zhang}, {B{\"o}hringer}, {Mellier}, {Soucail},
  \& {Forman}}]{zhang05}
{Zhang}, Y.-Y., {B{\"o}hringer}, H., {Mellier}, Y., {Soucail}, G., \& {Forman},
  W. 2005, \aap, 429, 85

\end{thebibliography}

\end{document}